\documentclass{ws-ijmpa}

\begin{document}

\markboth{Yue-Liang Wu and Yu-Feng Zhou}
{A two Higgs Bi-doublet Left-Right Model with Spontaneous $CP$ Violation}

%%%%%%%%%%%%%%%%%%%%% Publisher's Area please ignore %%%%%%%%%%%%%%%
%
\catchline{}{}{}{}{}
%
%%%%%%%%%%%%%%%%%%%%%%%%%%%%%%%%%%%%%%%%%%%%%%%%%%%%%%%%%%%%%%%%%%%%

\title{
A Two Higgs Bi-doublet Left-Right Model With Spontaneous $CP$ Violation
%
%INSTRUCTIONS FOR TYPESETTING 
%MANUSCRIPTS\footnote{For the title, try not to use more than 
%3 lines. Typeset the title in 10 pt roman, uppercase and 
%boldface.}
}

\author{Yue-Liang Wu}

\address{
Kavli Institute for Theoretical Physics China, Institute of Theoretical Physics, \\
Chinese Academy of Sciences, Beijing 100080, P.R.China
}

\author{Yu-Feng Zhou}

\address{
Korea Institute for Advanced Study, Seoul, 130-722, Korea
}

\maketitle

\begin{history}
%\received{Day Month Year}
%\revised{Day Month Year}
\end{history}

\begin{abstract}
We discuss a left-right symmetric model with two Higgs bi-doublet 
and spontaneous P and $CP$ violation. The
flavor changing neutral currents is suppressed by assuming
approximate global $U(1)$ family symmetry. We calculate the
constraints from neural $K$ meson mass difference $\Delta m_K$ and
demonstrate that a right-handed gauge boson $W_2$ contribution in
box-diagrams with mass around $600$ GeV is allowed due to a
negative interference with a light charged Higgs boson around
$150 \sim 300$ GeV. The $W_2$ contribution to $\epsilon_K$ is
suppressed from appropriate choice of additional $CP$ phases appearing
in the right-handed Cabbibo-Kobayashi-Maskawa(CKM) matrix. The model is
found fully consistent with $B^0$ mass difference and
the mixing-induced $CP$ asymmetry measurements. 

%\keywords{Keyword1; keyword2; keyword3.}
\end{abstract}

\ccode{PACS numbers: 12.60.Fr;13.25.Hw;11.30.Hv;}

Since the discovery of parity violation fifty years
ago\cite{LY-CSW}, the study of symmetry and symmetry breaking has
played a central role in particle physics.  With the hypothesis that
parity is a good symmetry at high energy scale, a minimal left-right
symmetric model was proposed based on the gauge group $SU(2)_L\times
SU(2)_R \times U(1)_{B-L}$\cite{LRM1-3} with one Higgs
bi-doublet. In such a model, parity violation is implemented through
spontaneous symmetry breaking of the right-handed gauge sector. It is
natural to require that in the left-right model, $CP$ violation is also a
consequence of spontaneous symmetry
breaking. However, a spontaneous
$CP$-violating left-right model with only one Higgs bi-doublet is
severely  constrained\cite{LRM5-7} by low energy
phenomenology as it generates too much flavor changing neutral
current(FCNC). For instance, the neutral kaon mass difference $\Delta
m_K$ requires that the right-handed gauge bosons must be very heavy
above 2 TeV and the lightest neutral Higgs boson must be above 10
TeV to suppress FCNC; the predicted CKM matrix
elements do not coincide with the B factory measurements; and the the
condition for the spontaneous $CP$ violation requires an unnatural fine
tuning of the Higgs potential.

Motivated by the general two-Higgs-doublet model(2HDM) as a model for spontaneous $CP$ violation,
we  extend the one Higgs bi-doublet left-right model to a
two Higgs bi-doublet model with spontaneous $CP$
violation \cite{Wu:2007kt}. The two Higgs bi-doublet fields are defined as
\begin{equation}\label{bidublet}
  \phi = \left(
    \begin{array}{cc}
      \phi_1^0  & \phi_1^+ \\
      \phi_2^- & \phi_2^0
    \end{array} \right),
  \quad
  \chi = \left(
    \begin{array}{cc}
      \chi_1^0  & \chi_1^+ \\
      \chi_2^- & \chi_2^0
    \end{array}
  \right)
  \quad :
  \left( 2,2,0 \right) .
\end{equation}
with most general Yukawa interaction for quarks
\begin{eqnarray}\label{yukava}
  \mathcal{L}_Y & = & - \sum\limits_{i,j}\bar{Q}_{iL}
  \left( (y_q)_{ij}\phi+ (\tilde{y}_q)_{ij}\tilde{\phi}  +
    (h_q)_{ij}\chi+ (\tilde{h}_q)_{ij}\tilde{\chi} \right) Q_{jR} ,
\end{eqnarray}
where $\tilde{\phi}(\tilde{\chi})  =  \tau_2\phi^{\ast}(\chi^{\ast})\tau_2$.
%
%Parity symmetry requires $ g_L =g_R \equiv g$. 
Parity and  $CP$
symmetry require that all the Yukawa matrices are real and symmetric.
 We also adopt the popular choice of introducing two
Higgs triplets $(\Delta_L$ $\sim (3,1,2)$, $\Delta_R$ $\sim (1,3,2)
)$ to break $SU(2)_L\otimes SU(2)_R \otimes U(1)_{B-L}$ down to
$U(1)_{em}$\cite{LRM1-3}

To suppress FCNC,
we shall follow the similar treatment in the general
2HDM\cite{WW} by considering the mechanism of
approximate global $U(1)$ family symmetry\cite{HW,WW}
\begin{align}
(u_i, d_i)\to e^{-i\theta_i} (u_i, d_i) ,
\end{align}
which is motivated by the approximate unity of the CKM matrix. As
an consequence,  $y$, $\tilde{y}$, $h$ and $\tilde{h}$ are nearly
diagonal matrices.
We argue that after the spontaneous symmetry breaking, the two Higgs
bi-doublet fields can have complex vacuum expectation values(VEVs)
which leads to  the following mass matrix for up-quarks
\begin{eqnarray}
M_u & = & y_q v_1 e^{i\delta_1} + \tilde{y}_q v_2
e^{-i\delta_2} + h_q w_1 e^{i\varphi_1} + \tilde{h}_q w_2
e^{-i\varphi_2} ,
%\nonumber\\
%M_d & = & y_q v_2 e^{i\delta_2} + \tilde{y}_q v_1
%e^{-i\delta_1} + h_q w_2 e^{i\varphi_2} + \tilde{h}_q w_1
%e^{-i\varphi_1} .
\end{eqnarray}
where $v_{1,2}$ and $w_{1,2}$ are VEVs for $\phi$ and $\chi$.
The down-quark mass matrix is obtained by replacing $1\leftrightarrow 2$.
It follows that the resultant
quark mixing matrices for left-handed and right-handed quarks are
of pseudo-manifest type before rephasing. 
After rephasing the left-handed CKM matrix $V^L$ to the standard form with
only one $CP$ phase, there are five $CP$ phases $\alpha_i,
(i=1,2,3)$ and $\beta_i, (i=1,2)$ appearing in the right-handed CKM matrix $V^R$
\begin{align*}
V^{R} & =\eta^{u}\left(\begin{array}{lll}
(V^{L}_{ud})^{*}e^{2i\alpha_{1}} & (V^{L}_{us})^{*}e^{i(\alpha_{1}+\alpha_{2}+\beta_{1})} & (V^{L}_{ub})^{*}e^{i(\alpha_{1}+\alpha_{3}+\beta_{1}+\beta_{2})}\\
(V^{L}_{cd})^{*}e^{i(\alpha_{1}+\alpha_{2}-\beta_{1})} & (V^{L}_{cs})^{*}e^{2i\alpha_{2}} & (V^{L}_{cb})^{*}e^{i(\alpha_{2}+\alpha_{3}+\beta_{2})}\\
(V^{L}_{td})^{*}e^{i(\alpha_{1}+\alpha_{3}-\beta_{1}-\beta_{2})}&
(V^{L}_{ts})^{*}e^{i(\alpha_{2}+\alpha_{3}-\beta_{2})} &
(V^{L}_{tb})^{*}e^{2i\alpha_{3}}\end{array}\right)\eta^{d} ,
\end{align*}
where  $\eta^{u,d}$ are quark mass sign matrices.

With enlarged Higgs sector, in this model there are two doubly charged
Higgs $H^{++}_i,(i=1,2)$, four singly charged Higgs particles $H^{+}_i,(i=1,\dots 4)$,
six neutral scalars $h^{0}_i,(i=1,\dots 6)$ and
four neural pseudo-scalars $A^{0}_i,(i=1,\dots 4)$.
%
%The doubly charged $H^{++}_i$  contribute  only to the leptonic sector
%such as lepton flavor violation processes\cite{Akeroyd:2006bb}, 
%whereas 
The
singly charged and neutral scalars may have significant effects on
mixings and $CP$ violation in quark sector.
For  simplicity, we shall work in a simple scenario
that only one charged Higgs (labeled as $H^+$) is light enough to
actively contribute to the box-diagrams in neutral meson mixing. Of particular, when the
VEVs satisfy the conditions $v_2\ll v_1$ and
$w_2\ll w_1$, many
features of this model are similar to the general 2HDM with
spontaneous $CP$ violation\cite{WW}. Therefore, we consider here
the 2HDM-like charged Higgs to be the lightest one, the
corresponding Yukawa interaction is  parametrized as follows
\begin{eqnarray*}
  \mathcal{L}_{C}&=&-(2\sqrt{2}G_{F})^{1/2}\bar{u}^i
  \left(
    \sqrt{m^u_{i} m^u_{k}} \ \xi^u_{ik} V^{L}_{kj}P_{L}-V^{L}_{ik}
    \sqrt{m^d_{k} m^d_{j}}\ \xi^d_{kj} P_{R}
  \right) d^j\ H^{+} +\mbox{H.c} .
\end{eqnarray*}
%Here we have used the Cheng-Sher parametrization\cite{CS} in the
%general 2HDM with $\xi^{u(d)}_{ij}$ the effective Yukawa coupling
%matrices in the physics basis after spontaneous symmetry breaking.
%Where $v =(\sqrt2 G_F)^{-1/2}$.
The small off-diagonal terms characterized in $\sqrt{m_{i}^q
  m_{j}^q}\ \xi^{q}_{ij}$ describe the breaking of the
global $U(1)$ family symmetry \cite{CS}. We denote
the diagonal elements as $\xi_c\equiv \xi^u_{22}$ and $\xi_t\equiv \xi^u_{33}$.

Let us discuss the low energy phenomenology.
In $K^0$ mixing, the smallness of $\epsilon_K=2.28\times 10^{-3}$ requires that 
the dominant $W_1 W_2$ box-diagrams with internal ($c,c$) quark must be
nearly real. This can be satisfied by imposing 
\begin{eqnarray}\label{phase}
\alpha_1-\alpha_2-\beta_1 \simeq 0 .
%\ \ \mbox{ or } \  \ \beta'_R
%\simeq -\beta'_L \simeq -\beta_L .
\end{eqnarray}
Since the $H^+$ contribution to $(c,c)$ quark loop is always real, and
the induced effective operator have different chirality,  the $H^+$
loop interferes always destructively with the $W_1 W_2$ loop in the
$CP$ conserving case of Eq.(\ref{phase}). This provides a possibility
of a cancellation, which may greatly reduce the mass
lower bounds for both $W_2$ and charged Higgs $H^+$.  Taking only the
dominant $H^\pm H^\pm$ loop contribution into account, we find that a
complete cancellation  requires
\begin{eqnarray}
  \eta^H_{cc} x_c |\xi_c|^4\frac{M_2^2}{m_H^2}
  &\simeq&-24
  \left[ \eta^{LR}_1 (\ln x_c+1)
    +\frac{\eta^{LR}_2}{4}\ln\frac{m_W^2}{M_2^2}
  \right]
  \frac{m_K^2}{(m_s+m_d)^2} \frac{B^S_K}{B_K} .
\end{eqnarray}
A numerical calculation including all the contributions is shown in
Fig.1a. Numerically, for $m_{H^+}\sim 150$GeV and Yukawa
coupling $\xi\sim 25$, the charged Higgs can compensate a opposite
contribution from a light $W_2$ at $M_2\sim 600$ GeV. The large Higgs
contribution relies on the fact that the $H^+H^+$ loop is proportional
to $|\xi_c|^4$ which grows rapidly with $|\xi_c|$ increasing.
\begin{figure}[htb]
\centerline{
\psfig{file=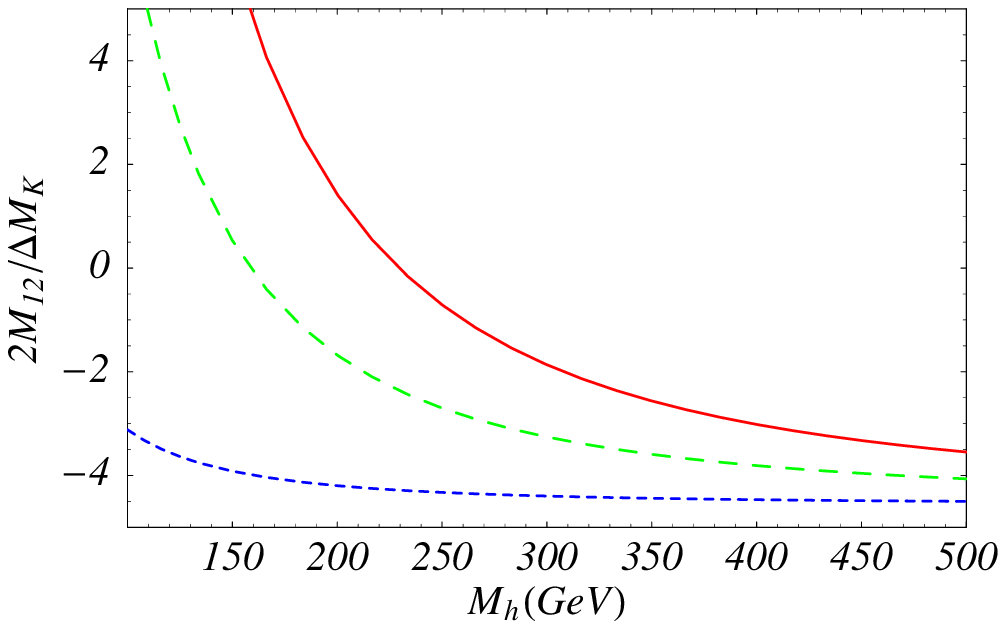,width=0.45\textwidth}
\psfig{file=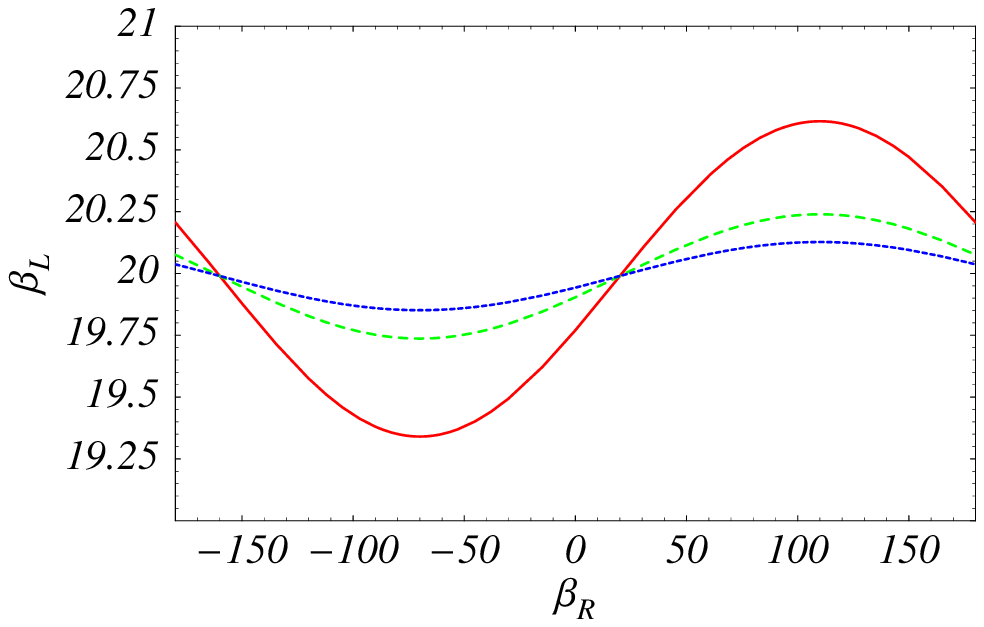,width=0.45\textwidth}
}
\vspace*{8pt}
\caption{(a)Left, sum of all loop contributions, including the SM
contribution to the $2M_{12}$ normalized to the measured $\Delta m_{K}$ with
$M_{2}=600$ GeV. Three curves corresponds to $|\xi_{c}|=$30(solid),
25( dashed) and 15 ( dotted ) respectively.
(b)Right,  Values of $\beta_{L}$ as a function of $\beta_{R}$ for
different $M_{2}$. Three curves correspond to $M_{2}=500$ GeV
(solid), 1000 GeV (dashed) and 1500 GeV(dotted) respectively.
\label{fig1}}
\end{figure}

The situation is quite different for $B$ meson system.  The $B^0$
meson mixing is dominated by internal $(t,t)$quark loop. Due to the
very weak mass-dependence of loop functions, the $W_1 W_2$ loop are only a
few percent of $W_1 W_1$ loop in the SM at $m_b$ scale, which greatly suppresses its
phenomenological significance in $B^0$ mixing. The charged Higgs
contribution can be suppressed for small $|\xi_t|$.
%
%Furthermore, the charged Higgs contribution is dominated by another
%Yukawa coupling $\xi_t$, and is suppressed for small $|\xi_t|$.
%
Within the Wolfenstein parametrization, we define $\beta_L \equiv
\mbox{arg}(V^{L*}_{td}V_{tb}^L )$ which is to a high precision, one of
the angles of the unitarity triangle. In the right-hand sector one can
define a similar quantity $\beta_R \equiv \mbox{arg}(V^{R*}_{td}
V_{tb}^R)$.
With the pollution from $W_1 W_{2}$ loop, the time-dependent decay
$B\to J/\psi K_{S}$ will only measures an effective phase angle
$\sin2\beta_{eff}$.  Using the measured experimental value of $\Delta
m_{B}$ and $\beta_{eff}$ one can obtain the value of $\beta_{L}$ as a
function of $\beta_{R}$ only.
%
% In the limit
% $m_{W}^{2}\ll M_{2}$, $\beta_{L}$ is close to $\beta_{eff}$ we have
% in a good approximation
% %
% \begin{equation}
% \tan2\beta_{L}\simeq
% \tan2\beta_{eff}\left[1-r\frac{\sin(\beta_{R}-\beta_{eff})}{2\sin4\beta_{eff}}\right] ,
% \end{equation}
% where $r$ is the ratio between $W_{1}W_{1}$ and $W_{1}W_{2}$ box
% diagrams. In Fig.(\ref{fig1}), 
%
In fig.1b, we plot the $\beta_L$ as a function of $\beta_R$ with
different values of $M_2$. One sees that for a light $W_2$ around 600
GeV, and $\beta_R$ varying from $-180^\circ$ to $180^\circ$, the
modification to $\beta_L$ is less than $2^\circ$. Similarly, the 
change to the  $|V^L_{td}|$, which is calculated from 
$B^0$ mass difference is also very small.
%
%
% Once the $\beta_L$
% is obtained, one can evaluate the matrix element $|V^L_{td}|$, it is
% found that the changes in $|V^L_{td}|$ is small for the whole range of
% $\beta_R$. Comparing with the global SM fit value of $|V^L_{td}|$ the
% modifications is within the $1\sigma$ error range.
%
Thus this model can  accommodate both the data of $\Delta m_B$
and $\sin2\beta_{J/\psi}$ with a light $W_2$, which is 
not possible to the left-right model with only one Higgs
bi-doublet.

% For the left-right model with only one Higgs
% bi-doublet, since both $\beta_L$ and $\beta_R$ are calculable quantities
% which depends only on the quark masses and ratios of VEVs, there is
% little room to meet the $CP$ violation in both $K$ and $B$ system. Due to
% the suppression of $CP$ phase form $\epsilon_K$, the predicted
% $\sin2\beta_{J/\psi}$ has to be small and can not excess $0.1$.

To summarize, we have investigated a
general left-right symmetric model with two Higgs bi-doublets. This
simple extension evades the stringent constraints from $K$ and 
$B$ meson mixing, and lowers the allowed mass of right-handed gauge boson
closing to the current direct experimental search bound.
%  The FCNC is suppressed by the mechanism of
%approximate global $U(1)$ family symmetry.  
Compared with the general 2HDM, this model has even more sources of $CP$
violation in gauge sector, which may have interesting implications in 
lower energy phenomenology such as $\epsilon'/\epsilon$\cite{WW} and  
puzzles in charmless $B$ decays\cite{BdecayNP}.  
The predicted new physics particles can be directly
searched in upcoming LHC and future ILC experiments.

%\begin{thebibliography}{000} %for 3 digits
%\begin{thebibliography}{00}  %for 2 digits

\end{document}